\documentstyle[epsfig]{article}
\begin{document}
\voffset .3in
\bibliographystyle{unsrt}    

\def\ie{{\it i.e.}}
\def\etal{{\it et al.}}
\def\9{\phantom 0}     
\def\epem{e^+e^-}
\def\lumi{cm$^{-2}$s$^{-1}$}
\def\bold#1{\setbox0=\hbox{$#1$}%
     \kern-.025em\copy0\kern-\wd0
     \kern.05em\copy0\kern-\wd0
     \kern-.025em\raise.0433em\box0 }
\def\lsim{\stackrel{<}{{}_\sim}}
\def\gsim{\stackrel{>}{{}_\sim}}

\def\be{\begin{equation}}
\def\ee{\end{equation}}
\def\bea{\begin{eqnarray}}
\def\eea{\end{eqnarray}}
\def\CPbar{\hbox{{\rm CP}\hskip-1.80em{/}}}
\def\NPB#1#2#3{{\sl Nucl.~Phys.} {\bf{B#1}} (19#2) #3}
\def\PLB#1#2#3{{\sl Phys.~Lett.} {\bf{B#1}} (19#2) #3}
\def\PRD#1#2#3{{\sl Phys.~Rev.} {\bf{D#1}} (19#2) #3}
\def\PRL#1#2#3{{\sl Phys.~Rev.~Lett.} {\bf{#1}} (19#2) #3}
\def\ZPC#1#2#3{{\sl Z.~Phys.} {\bf C#1} (19#2) #3}
\def\PTP#1#2#3{{\sl Prog.~Theor.~Phys.} {\bf#1}  (19#2) #3}
\def\MPL#1#2#3{{\sl Mod.~Phys.~Lett.} {\bf#1} (19#2) #3}
\def\PR#1#2#3{{\sl Phys.~Rep.} {\bf#1} (19#2) #3}
\def\RMP#1#2#3{{\sl Rev.~Mod.~Phys.} {\bf#1} (19#2) #3}
\def\HPA#1#2#3{{\sl Helv.~Phys.~Acta} {\bf#1} (19#2) #3}
\relax
\topmargin-2.5cm
\begin{flushright}
\large {CERN-TH/96-07\\
SCIPP 96/05 \\
hep--ph/9601330} \\
\end{flushright}
\vskip 1.0in
\begin{center}
{\LARGE\bf Recent Refinements in Higgs Physics}\\
\vskip .5in
{\large 
{\bf Howard E. Haber} \\
\vskip .35in
CERN, TH-Division, \\
CH--1211 Geneva 23, Switzerland \\[2pt]
{\rm and}\\[2pt]
Santa Cruz Institute for Particle Physics, \\
University of California, Santa Cruz, CA 94064  USA}\\
\end{center}
\vskip1.75cm
\begin{center}
{\large \bf Abstract}
\end{center}
\begin{quote}
{\large
Recent refinements of the phenomenology of Higgs bosons in the
Standard Model and the Minimal Supersymmetric Standard Model 
are reviewed.}
\end{quote}
\vfill
\begin{center}
{\large Invited Talk at the \\
1995 International Europhysics Conference on High Energy Physics,\\
Brussels, Belgium, 27 July--2 August, 1995}\\
\end{center}
\vskip 1.5cm
\begin{flushleft}
{\large CERN-TH/96-07\\
January, 1996}\\
\end{flushleft}
\setcounter{page}{0}
\newpage
%
\makeatletter
\input{wstwocl.sty}
\newcount\@tempcntc
\def\@citex[#1]#2{\if@filesw\immediate\write\@auxout{\string\citation{#2}}\fi
  \@tempcnta\z@\@tempcntb\m@ne\def\@citea{}\@cite{\@for\@citeb:=#2\do
    {\@ifundefined
       {b@\@citeb}{\@citeo\@tempcntb\m@ne\@citea\def\@citea{,}{\bf ?}\@warning
       {Citation `\@citeb' on page \thepage \space undefined}}%
    {\setbox\z@\hbox{\global\@tempcntc0\csname b@\@citeb\endcsname\relax}%
     \ifnum\@tempcntc=\z@ \@citeo\@tempcntb\m@ne
       \@citea\def\@citea{,}\hbox{\csname b@\@citeb\endcsname}%
     \else
      \advance\@tempcntb\@ne
      \ifnum\@tempcntb=\@tempcntc
      \else\advance\@tempcntb\m@ne\@citeo
      \@tempcnta\@tempcntc\@tempcntb\@tempcntc\fi\fi}}\@citeo}{#1}}
\def\@citeo{\ifnum\@tempcnta>\@tempcntb\else\@citea\def\@citea{,}%
  \ifnum\@tempcnta=\@tempcntb\the\@tempcnta\else
   {\advance\@tempcnta\@ne\ifnum\@tempcnta=\@tempcntb \else \def\@citea{-}\fi
    \advance\@tempcnta\m@ne\the\@tempcnta\@citea\the\@tempcntb}\fi\fi}
\makeatother
\setcounter{secnumdepth}{2} 


\title{RECENT REFINEMENTS IN HIGGS PHYSICS}

\firstauthors{Howard E. Haber}

\firstaddress{CERN, TH-Division, CH-1211 Geneva 23, Switzerland}

\secondaddress{Santa Cruz Institute for Particle Physics,
University of California, Santa Cruz, CA, 95064, USA}

\twocolumn[\maketitle
\abstracts{ 
Recent refinements of the phenomenology of Higgs bosons in the
Standard Model and the Minimal Supersymmetric Standard Model are reviewed.}
]

\pagestyle{plain}
\section{Introduction}

In this brief review, I will summarize a few of the advances in
Higgs phenomenology during 1995.  Space limitations
allow me to present only a bare outline of the talk presented at
the EPS meeting in Brussels, along with the attendant references. 
As a result, this review is necessarily incomplete, and I apologize in
advance to authors of relevant work that has been omitted. 

\section{The Standard Model Higgs Boson is a Contradiction in Terms}

Phenomenologists and experimentalists who plan the Higgs searches at
future colliders spend much effort in designing a search for the
Standard Model Higgs boson.  However, the Standard Model Higgs boson
is a meaningless term unless additional information is provided.  This
is because the Standard Model itself cannot be a fundamental theory of
particle interactions.  It must break down once the energy is raised
beyond some critical scale $\Lambda$.  What is the value of $\Lambda$?
Of course, this is unknown at present.  $\Lambda$ can lie anywhere
between a few hundred GeV and the Planck scale ($M_{\rm PL}$).

Theorists who discuss the phenomenology of the Standard Model usually
do not need to know the value of $\Lambda$.  At energy scales below
$\Lambda$, the physics beyond the Standard Model generally decouples,
leaving a low-energy effective theory which looks almost exactly like
the Standard Model.  However, the Higgs boson presents a potential
opportunity to probe $\Lambda$.  The stability of the Higgs potential
places non-trivial constraints on the Higgs mass, due to the large
value of the top quark mass.  (More refined limits require only a
metastable potential with a lifetime that is long
compared to the age of the universe.)  Recent computations of
Refs.~1 and 2 show for example that if $\Lambda=M_{\rm PL}$,
then for $m_t=175$~GeV the Higgs mass must be
larger than about 120~GeV.  

Does this mean that if a Higgs boson mass
of 100~GeV is discovered then the Standard Model Higgs boson is ruled
out?  The answer is yes, only if the phrase ``the Standard Model
Higgs boson'' implies that $\Lambda=M_{\rm PL}$.
For me, this is too narrow a definition.  I would prefer
to say that if a 100~GeV Higgs boson were discovered, then new physics
beyond the Standard Model must enter at or below an energy scale of
$\Lambda\simeq 1000$~TeV (based on the graphs presented in
Ref.~2).  Of course, in this case, if all the new physics were
confined to lie in the vicinity of 1000~TeV, then LHC phenomenology would
find no deviations from the Standard Model.  Thus, physicists who
plan searches for the Standard Model Higgs boson are not wasting their
time.  In particular, even if $\Lambda$ is rather close to the TeV
scale, one would expect the lightest Higgs boson to retain all the
properties of the so-called Standard Model Higgs boson.

To reiterate, the Standard Model Higgs boson is a sensible concept
only if you specify the value of the energy scale $\Lambda$ at which
the Standard Model breaks down.
Now that we can all agree on the meaning of ``the Standard Model Higgs
boson'', consider recent refinements in its phenomenology.  Here, I
would like to refer to two interesting directions.  

First, if one
assumes the validity of the Standard Model for low-energy physics
(below the TeV scale), then one can test this theory by confronting it
with the precision electroweak data.  In addition to testing the
Standard Model, one has the possibility of constraining the value of
the Higgs mass, which enters through the radiative corrections to the $Z$
and $W$ boson self-energies.  Combining the most recent LEP and SLC
electroweak results~\cite{r4} with the recent top-quark mass measurement
at the Tevatron,~\cite{r5} a weak preference is
found~\cite{r4,r4a} for a light Higgs boson mass of order $m_Z$.
One must take this result with a large grain of salt, since the
overall chi-square of the Standard Model fit is not good (of order 2
per degree of freedom).  Nevertheless, it does suggest the potential
of future precision measurements for placing interesting constraints on
the Standard Model Higgs mass.  

Second, a number of two-loop computations of Higgs boson processes
have recently been completed.  Among them are an ${\cal
O}(\alpha_s^3)$ calculation of $h^0\to gg$ and ${\cal O}(\alpha_s^2,
\alpha_s G_F m_t^2, G_F^2 m_t^4)$ terms in $h^0\to b\bar b$.
See Ref.~6 for details.
Together with the recent computation of ``$K$''-factors in 
$pp\to h^0+X$,\cite{kfactors}
one now has both improved production cross-section and branching ratio  
calculations, leading to more accurate Higgs boson 
phenomenology at both the LHC and future $e^+e^-$ colliders.

\section{The Radiatively-Corrected MSSM Higgs Mass}

If the minimal supersymmetric extension of the Standard Model (MSSM)
is correct, then we should identify the scale $\Lambda$ at which the
Standard Model breaks down as the scale of low-energy supersymmetry
breaking.  In models of low-energy supersymmetry, $\Lambda$ is
presumed to lie between $m_Z$ and about 1~TeV.
The mass of the light CP-even neutral Higgs boson, $h^0$, in the MSSM
can be calculated to arbitrary accuracy in terms of two parameters of
the Higgs sector, $m_{A^0}$ and $\tan\beta$,~\cite{hhg} and other MSSM
soft-supersymmetry-breaking parameters that affect the Higgs mass
through virtual loops.~\cite{hhprl}  If the
scale of supersymmetry breaking is much larger than $m_Z$, then large
logarithmic terms arise in the perturbation expansion.  These large
logarithms can be resummed using renormalization group (RG) methods.

The formula for the full one-loop radiative corrected Higgs mass is
very complicated.~\cite{honeloop}  Moreover, the computation of the
RG-improved one-loop
corrections requires numerical integration of a coupled set of
RG equations.~\cite{llog} (The dominant two-loop next-to-leading 
logarithmic results are also known.~\cite{hempfhoang})  
Although this program has been
carried out in the literature, the procedure is unwieldy
and not easily amenable to large-scale Monte-Carlo analyses.  Below,
we summarize a very simple procedure for accurately approximating $m_{h^0}$.  
The method can be easily implemented, and incorporates both the
leading one-loop and two-loop effects and the RG-improvement.
Although the method is conceptually simple, complications arise when
supersymmetric thresholds are fully taken into account.  The details can
be found in Ref.~13, along with other references to the original
literature.  Complementary work can be found in Ref.~14.
In the limited space alloted here, only the simplest version
of our method is outlined.

The dominant radiative corrections to $m_{h^0}$ arise from an incomplete
cancelation of the virtual top-quark and top-squark loops.
The two top-squark masses 
($M_{\widetilde t_1}$ and $M_{\widetilde t_2}$) are obtained by
diagonalizing a $2\times 2$ top-squark squared-mass matrix; the
off-diagonal elements of this matrix are denoted by $m_t X_t$ 
(where $X_t\equiv A_t-\mu\cot\!\beta$).
We also assume that 
$M_{\widetilde t_1}$, $M_{\widetilde t_2}$, $m_{A^0}\gg m_Z$.
This case is particularly useful since the upper bound
for $m_{h^0}$ (at fixed $\tan\beta$) arises precisely in this limit.
The leading terms in the one-loop approximation to $m_{h^0}^2$ are given by
\begin{equation} \label{hlmass}
m_{h^0}^2= m_Z^2\cos^2 2\beta+(\Delta m_{h^0}^2)_{\rm
1LL}(m_t)+(\Delta m_{h^0}^2)_{\rm mix}(m_t)\,,
\end{equation}
where
\begin{equation}
(\Delta m_{h^0}^2)_{\rm 1LL}= {3g_2^2m_t^4\over 8\pi^2m_W^2}
 \ln\left({M_{\widetilde t_1}M_{\widetilde t_2}
\over m_t^2}\right)\left[1+{\cal O}\left(
{m_W^2\over m_t^2}\right)\right]\,,
\label{mhlonell}
\end{equation}
and
\begin{eqnarray}
(\Delta m_{h^0}^2)_{\rm mix} & = & {3g_2^2m_t^4X_t^2\over16\pi^2m_W^2}
\Biggl[2h(M_{\widetilde t_1}^2,M_{\widetilde t_2}^2)
\nonumber \\
&&+X_t^2 g(M_{\widetilde t_1}^2,M_{\widetilde t_2}^2)\Biggr]
\left[1+{\cal O}\left({m_W^2\over m_t^2}\right)\right]\,.
\label{mhlmix}
\end{eqnarray}
In eq.~(\ref{mhlmix}), the functions $g$ and $h$ are given by:
\begin{eqnarray}
g(a,b)&\equiv& {1\over (a-b)^2}\left[2-{a+b\over a-b}\ln\left({a\over b}
\right)\right]\,,\label{gdef} \\
h(a,b)&=&{1\over a-b}\ln\left({a\over b}\right)\,.\label{hdef}
\end{eqnarray}

The notation used in eq.~(\ref{hlmass}) emphasizes the $m_t$ dependence
of $(\Delta m_{h^0}^2)_{\rm 1LL}$ and $(\Delta m_{h^0}^2)_{\rm mix}$.
Subdominant terms not shown explicitly in eqs.~(\ref{mhlonell})
and (\ref{mhlmix}) can be found in ref.~13.
Among the next-to-leading two-loop corrections, the most important 
effect can be incorporated 
by replacing $m_t$ in the formulae above by the running top quark mass 
evaluated at $m_t$.  Explicitly, $m_t(m_t)$ is given in terms of the
pole mass $m_t^{\rm pole}$ by
\begin{equation} \label{runmass}
m_t(m_t)= m_t^{\rm pole}\,\left(1 - \frac{4\alpha_s}{3\pi} +
    \frac{\alpha_t}{2\pi}\right)\approx 0.966\,m_t^{\rm pole}\,,
\end{equation}
where $\alpha_t\equiv h_t^2/4\pi$ and the Higgs-top quark Yukawa
coupling is $h_t=gm_t/\sqrt{2}m_W$.  (Since by assumption
$m_{A^0}\gg m_Z$, one should identify $h_t$ as the Yukawa coupling of the
low-energy effective one-doublet Higgs sector.  For details
appropriate to the case of $m_{A^0}\sim{\cal O}(m_Z)$, see
Ref.~13.)   Due to the leading $m_t^4$ behavior of the
one-loop corrections, the replacement of $m_t=m_t^{\rm pole}$ in
eqs.~(\ref{mhlonell}) and (\ref{mhlmix}) by $m_t(m_t)$
is numerically important,
leading to a significant reduction in the predicted value of $m_{h^0}$.

\begin{figure*}
\centerline{
\epsfig{figure=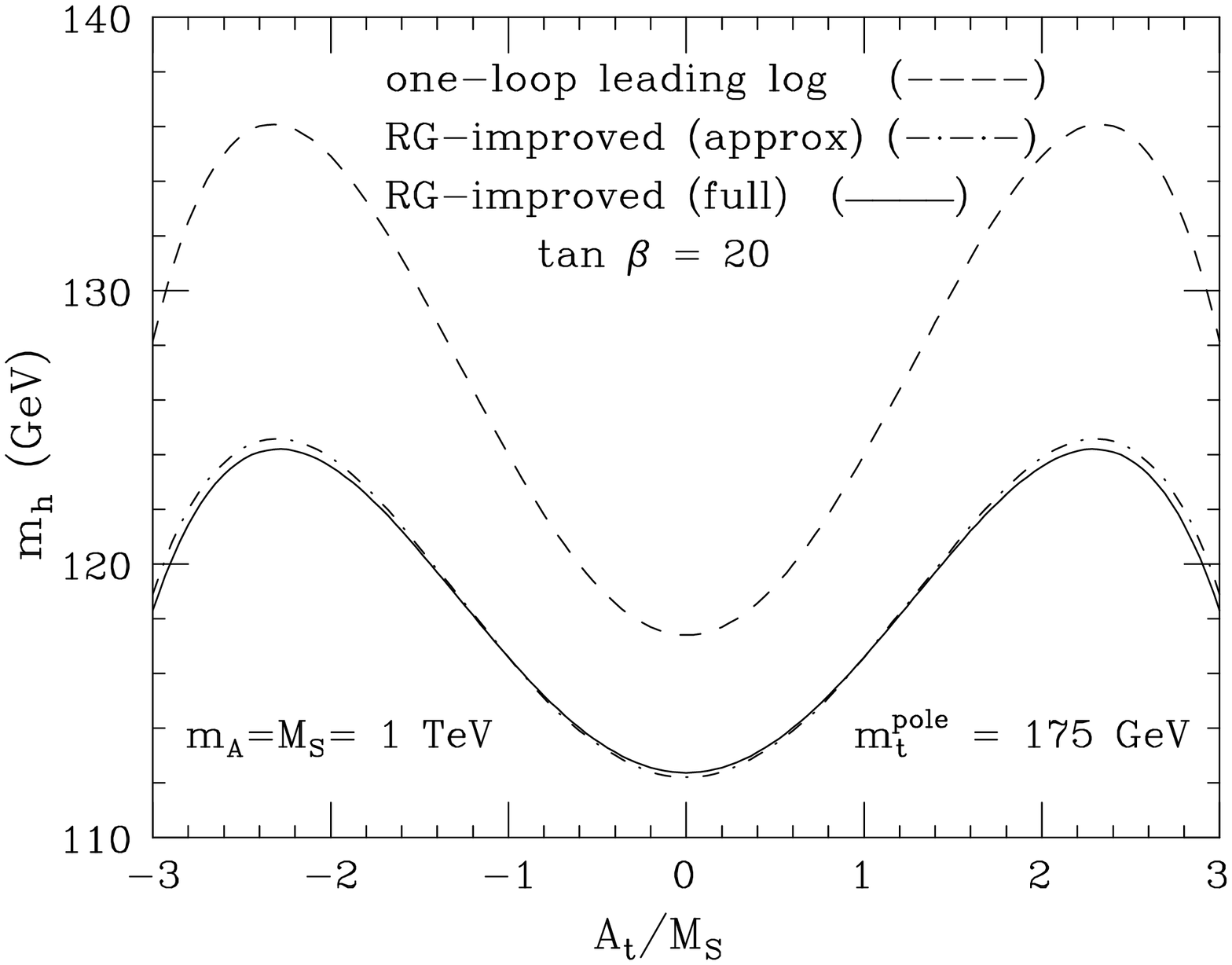,width=8.5cm,angle=90}
\hfill
\epsfig{figure=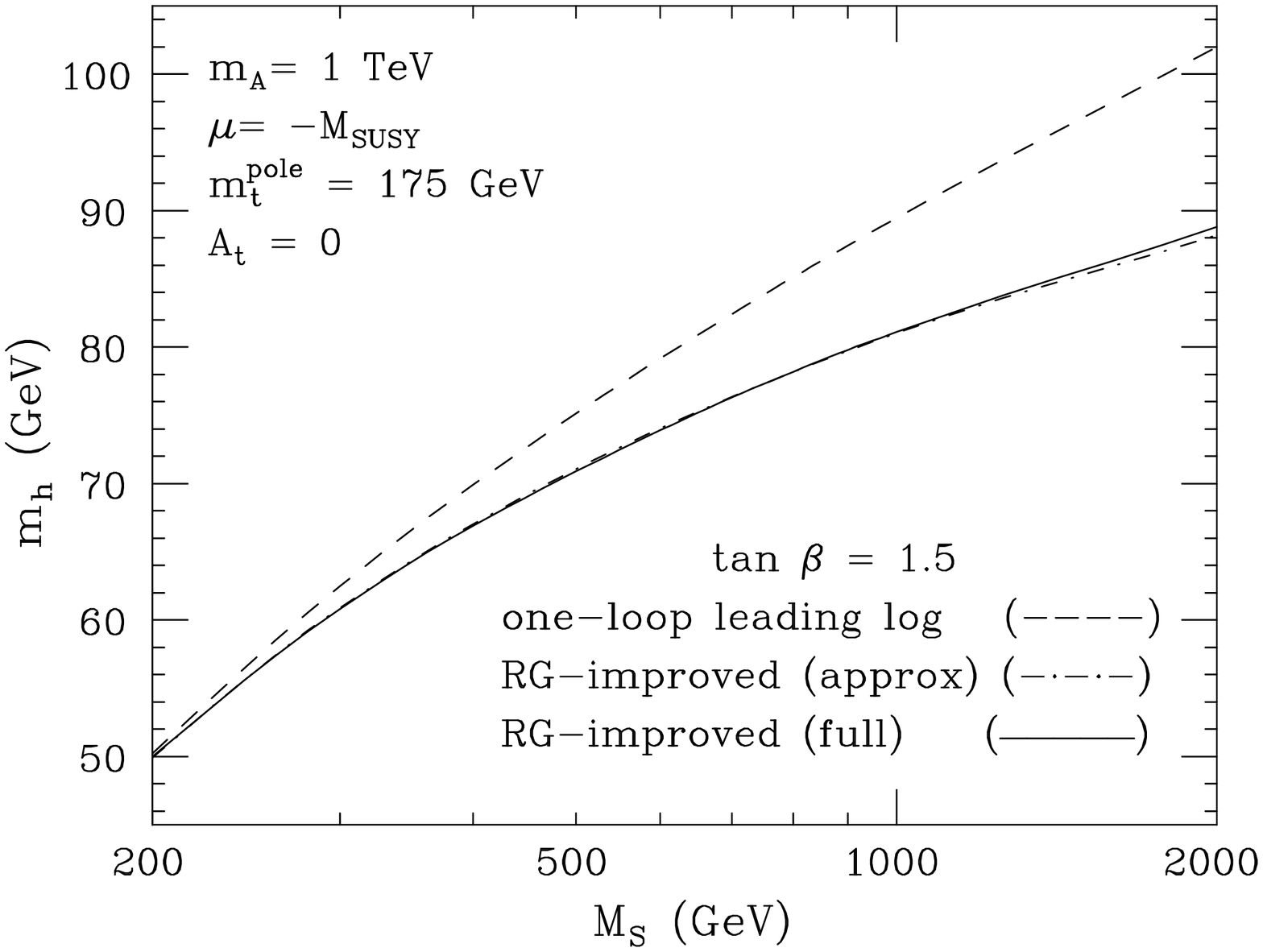,height=6.5cm,angle=90}
}
\begin{minipage}{8.5cm}       
\caption{\it
The radiatively corrected light CP-even Higgs mass is plotted
as a function of $A_t/M_S$ for $\tan\beta=20$.
The one-loop leading logarithmic
computation is compared with the RG-improved result which was obtained
by numerical analysis and by using the simple analytic result given in 
eq.~(\protect\ref{hlmassimproved}).  
$M_S$ characterizes the
scale of supersymmetry breaking and can be regarded (approximately)
as a common supersymmetric scalar mass.}
\label{fig1}
\vspace*{1pc}
\end{minipage}
%
\hfill
\begin{minipage}{8.5cm}       
\caption{\it
The radiatively corrected light CP-even Higgs mass is plotted
as a function of $M_S$ for $\tan\beta=1.5$.  For further details, see
the caption to Figure 1.}
\label{fig2}
\end{minipage}
\end{figure*}

We now proceed to sum the leading logarithmic terms to all orders in
perturbation theory via RG-improvement.  As noted above, this
requires extensive numerical analysis.  Nevertheless, we have found a
remarkably simple analytic formula that incorporates the dominant
effects of the RG-improvement.  
We already noted that replacing the
pole mass $m_t^{\rm pole}$ with the running mass $m_t(m_t)$ has the effect of
including the dominant part of the ${\cal O}(m_t^2\alpha_t^2)$ and 
${\cal O}(m_t^2\alpha_t\alpha_s)$ next-to-leading logarithmic
contributions to $m_{h^0}^2$.  We find that the two-loop leading
logarithmic contributions to $m_{h^0}^2$ can be incorporated by replacing
the running top quark mass with $m_t$ evaluated at an appropriately chosen
scale.  For $M_{\tilde t_1}\approx M_{\tilde t_2}\equiv
M_{\tilde t}$, our result is:
\begin{eqnarray} \label{hlmassimproved}
m_{h^0}^2 &=& m_Z^2\cos^2 2\beta+(\Delta m_{h^0}^2)_{\rm
1LL}(m_t(\mu_t)) \nonumber \\
&&\qquad+ (\Delta m_{h^0}^2)_{\rm mix}(m_t(M_{\tilde t}))\,,
\end{eqnarray}
where $\mu_t\equiv\sqrt{m_t M_{\tilde t}}$,
and the running top-quark mass is given by
\begin{equation}
m_t(\mu) = m_t(m_t)\left[
1-\left({\alpha_s\over\pi}-{3\alpha_t\over 16\pi}\right)\,
\ln\left({\mu^2\over m_t^2}\right)\right]\,.
\label{mtrun}
\end{equation}
All couplings on the right hand side of eq.~(\ref{mtrun})
are evaluated at $m_t$.  In
our numerical work, we have verified that this prescription reproduces the
full RG-improved Higgs mass to within 2 GeV for top-squark masses of
2~TeV or below.  Figures~\ref{fig1} and \ref{fig2}
exhibit two graphs to support this
claim.  Further details can be found in Ref.~13.

\section{Higgs Searches at Future Colliders}

1995 was a year of re-assessment.  In the United States, the Division
of Particles and Fields published a Long Range Planning Study to
assess the future of the US particle physics program in light of the
demise of the SSC.  One working group of this study focused on
electroweak symmetry breaking and physics at the TeV scale.  One of
the tasks of this working group was to provide an in depth survey of
Higgs searches at future collider facilities.  Their work can be found
in Ref.~15.  In Europe, 1995 was the year of the LEP-200
Study.  The Higgs Boson Working group played a major role in making
the case for upgrading the LEP collider to the highest possible
energy.  It now appears that LEP will be upgraded to a center-of-mass
energy of 192~GeV during the next few years.  This will permit the
discovery of a Higgs boson up to around 95 GeV.  This is very exciting
for proponents of the MSSM if $\tan\beta$ is small [see
Figure~\ref{fig2}].  
The LEP-200 Higgs Working group considered in
detail the discovery reach of the upgraded LEP collider for the
Standard Model Higgs boson and the MSSM Higgs boson (a few non-minimal
approaches were also surveyed).  The results of this work can be found
in Ref.~16.  

\section{Conclusions}

This past year has seen a number of theoretical and phenomenological
refinements of Higgs boson properties.  The most comprehensive
assessments of the Higgs searches at future colliders have been
presented.  Meanwhile, we wait for experiments at LEP-2 and/or the LHC
to shed light on the origin of the dynamics that is responsible for
electroweak symmetry breaking.

\section*{Acknowledgments}

The work of Section 3 is based on a collaboration with Andre Hoang and
Ralf Hempfling.  
This work was supported in part by the U.S. Department of Energy.

\section*{References}
 
\clearpage
 
\end{document}